\documentclass[sigconf]{acmart}

\usepackage{booktabs} 
\usepackage{url}  
\usepackage{graphicx}  
\usepackage{color}
\usepackage{hyperref}
\usepackage{tikz}
\usepackage{pgfplots}
\usepackage{float}
\usepackage{amsmath,amssymb}
\usepackage{color}
\usepackage{booktabs}
\usepackage{multirow}
\usepackage{bbm}
\usepackage[linesnumbered,ruled]{algorithm2e}
\usepackage{caption}
\usepackage{subcaption}

\DeclareMathOperator*{\E}{\mathbb{E}}

\newcommand{\bL}{\boldsymbol{L}}

\newcommand{\bxi}{\boldsymbol{\xi}}
\newcommand{\bepsilon}{\boldsymbol{\epsilon}}
\newcommand{\btheta}{\boldsymbol{\theta}}

\newcommand{\bmu}{\boldsymbol{\mu}}
\newcommand{\bSigma}{\boldsymbol{\Sigma}}

\newcommand{\bPsi}{\boldsymbol{\Psi}}

\newcommand{\bomega}{\boldsymbol{\omega}}

\newcommand{\bI}{\boldsymbol{I}}
\newcommand{\bzero}{\boldsymbol{0}}
\newcommand{\brho}{\boldsymbol{\rho}}
\newcommand{\bC}{\boldsymbol{C}}
\newcommand{\bv}{\boldsymbol{v}}

\setcopyright{rightsretained}

\newcommand\T{\rule{0pt}{2.9ex}}       
\newcommand\B{\rule[-1.2ex]{0pt}{0pt}} 

\begin{document}
\title{Latent Variable Session-Based Recommendation }

\author{Stephen Bonner}
\orcid{1234-5678-9012}
\affiliation{%
   \institution{Department of Computer Science, Durham University}
   \city{Durham}
 }
\email{s.a.r.bonner@durham.ac.uk}

\author{David Rohde}
\affiliation{%
  \institution{Criteo Research}
  \city{Paris}
}
\email{d.rohde@criteo.com}


\begin{abstract}

Session based recommendation provides an attractive alternative to the traditional feature engineering approach to recommendation.  Feature engineering approaches require hand tuned features of the user's history to be created to produce a context vector.  In contrast a session based approach is able to dynamically model the user's state as they act.
We present a probabilistic framework for session based recommendation.  A latent variable for the user state is updated as the user views more items and we learn more about their interests.  The latent variable model is conceptually simple and elegant; yet requires sophisticated computational technique to approximate the integral over the latent variable. We provide computational solutions using both the re-parameterization trick and also using the Bouchard bound for the softmax function, we further explore employing a variational auto-encoder and a variational Expectation-Maximization algorithm for tightening the variational bound.  The model performs well against a number of baselines.  The intuitive nature of the model allows an elegant formulation combining correlations between items and their popularity and that sheds light on other popular recommendation methods.  An attractive feature of the latent variable approach is that, as the user continues to act, the posterior on the user's state tightens reflecting the recommender system's increased knowledge about that user.

\end{abstract}

\acmDOI{}

\acmISBN{}

\acmConference[]{}{}{}
\acmYear{2019}
\copyrightyear{2019}

\acmArticle{4}
\acmPrice{15.00}



\maketitle

\section{Introduction}

A traditional approach to building a recommender system is to use feature engineering techniques in order to summarize a user's history into a feature vector of fixed dimension, which enables machine learning algorithms to be applied in order to do next item predictions or to model the outcome of recommendations. Feature engineering of the variable dimension user history is often quite compromised, for example the simple heuristic of looking at the most recent item is often employed.  Session based recommendation represents a significant step forward where instead of producing a feature vector there is a representation of the recommender system's state of knowledge about the users interests at a certain point in time.

Session based models require the temporal and sequential features of the user behavior to be modeled.  In this approach, rather than feature engineering being used to build a model, a user's state is dynamically updated as the user acts and responds to recommendations.  This has traditionally been approached in the recommender system community chiefly using Recurrent Neural Network (RNN) based approaches \cite{hidasi2018recurrent} \cite{quadrana2018sequence} \cite{zolna2017user} \cite{quadrana2017personalizing} \cite{smirnova2017contextual} \cite{quadrana2018sequence} \cite{tan2016improved} for other approaches see \cite{ying2018sequential} \cite{kang2018self} \cite{shani2005mdp}. RNNs are a powerful temporal models which can model subtle user dynamics, for example users may visit certain items with higher probability in a certain order.

There are two reasons that the recommender systems representation of a user may change in time, the first is the temporal nature of the user's interests e.g. they are in market for a product until they buy it and then they no longer are, secondly the recommender systems understanding of the user will improve as the user performs more actions and reveals more of their implicit interests.

Employing session based recommendation is also an important step towards modeling long term rewards, such as sales. This is due to the fact that a short term reward (e.g. a click) might be reasonable using a within subject study design \cite{greenwald1976within} but a between subject (i.e. session based) is needed for long term rewards.

Our primary contributions in this paper are as follows:
\begin{itemize}
\item
  We demonstrate how much of the power of embedding methods can be recovered through the use of a low-rank multivariate Gaussian latent variable model, transformed to a categorical output via a softmax. Our framework is semi-Bayesian integrating the latent variable with associated computational challenges, despite the model's apparent simplicity and elegance.
  \item 
  We provide an elegant way to convert a user history, containing a variable number of item views, into a fixed dimensional representation of the user's history.
  \item
We derive an analytical variational bound for our model and show how to train embeddings using a variational auto-encoder.  We also show that the model can be trained without the bound using the re-parameterization trick.
\item
We show that the variational auto-encoder can be used at prediction time to produce a user representation, we also derive a variational EM algorithm for the same purpose.
\item
We show that the method performs well on both synthetic and real-world data.
\end{itemize}


In Section~\ref{back} we introduce the latent variable model and show that it has intuitively pleasing properties.  In Section~\ref{sec:lit-review} we review relevant literature.  In Section~\ref{sec:aprox-inf} we explain our computational framework for performing approximate inference.  In Section~\ref{sec:experimental-setup} we outline the experimental setup and in Section~\ref{sec:results}, finally Section~\ref{sec:conclusion} makes some concluding remarks.

\section{Background}
\label{back}
\subsection{A Latent Variable Model For Item Views}

The model we introduce in this paper is that item views in a session are explained by a session level latent variable.  The model may be either viewed as a probabilistic matrix factorization model or as a co-variance estimation of a low rank Multivariate normal.  Perhaps surprisingly when combined with computational machinery to marginalize the latent variable, this simple model is able to reproduce many interesting features of a session based recommender system.

\begin{table}
\begin{tabular}{c l r}
  \toprule
  \textbf{Symbol} & \textbf{Dimension} & \textbf{Description} \\ 
  \midrule
  \midrule
  $u$ & Scalar & A given user's id. \T \\
  $t$ & Scalar & sequential time. \\
  $P$ & Scalar & Total number of products. \\
  $K$ & Scalar & The size of the embedding. \\
  $v_{u,t}$ & Scalar & Product id for user $u$ at time $t$.\\
  $\bomega_u$ & $K \times 1$& A given user's state.\\
  $\bPsi$ & $P \times K$ & Product embedding matrix.\\
  $\bPsi_v$ & $1 \times K$ & Product embedding for $v$.\\
  $\bmu_q$ & $K \times 1$ & The mean of $\bomega_u$.\\
  $\bSigma_q$ & $K \times K$& The covariance of $\bomega_u$.\\
  $\brho$ & $P \times 1$ & Item popularity shift.\\
  $T_u$ & Scalar & Session length for $u$. \B \\
  \bottomrule
  
  \end{tabular}
  \caption{Notations and Definitions} 
  \vskip -15pt
  \label{tab:notation}
\end{table}

Our model describes a generative process for the types of products that user's co-view in sessions. Throughout this paper, we will make use of the notation introduced in Table \ref{tab:notation}. We use $u$ to denote a user or a session, we use $t$ time to denote sequential time and $v$ to denote which product they viewed from $1$ to $P$ where $P$ is the number of products, the user's interest is described by a $K$ dimensional latent variable $\bomega_{u}$ which can be interpreted as the user's interest in $K$ topics. The session length of user $u$ is given by $T_u$.  We then assume the following generative process for the views in each session:

\[
\bomega_u \sim \mathcal{N}(\bzero_K, \bI_K)
\]

\[
v_{u,1}, .., v_{u,T_u} \sim {\rm categorical}({\rm softmax} (\bPsi \bomega_u + \brho) ) 
\]

For the moment we assume that $\bPsi$ and $\brho$ have already been estimated, we defer the topic of estimation to later sections.  In production we have observed a user's online viewing history  $v_{u,1}, .., v_{u,T_u}$ and we would like to produce a representation of the user's interests.  Our proposal is to use Bayesian inference in order to infer $p(\bomega|v_{u,1}, .., v_{u,T_u}, \bPsi, \brho)$ as a representation of their interests.  This representation of interests can then be used as a feature for training a recommender system on so called \emph{``bandit feedback''} i.e. logs of the recommender system itself (this is distinct from the user history).  The above is a \emph{``matrix factorization''} view of our model, however one could also view it as a \emph{``covariance-estimation''} task:

\[
\btheta_u \sim \mathcal{N}(\brho, \bPsi \bPsi^T),
\]

\[
v_{u,1}, .., v_{u,T_u} \sim {\rm categorical}({\rm softmax} (\btheta) ). 
\]

\noindent
We neglect the mathematical complications due to the covariance matrix: $\bPsi \bPsi^T$
 (usually) being low-rank and the fact that $\btheta_u$ formally has no density.


\subsection{Case Study}

Before going into the details of the model and related computational material we demonstrate on a simple case study that this model is deceptively powerful in producing an update-able representation of a user's interests.

Imagine we have a recommender system that has just seven products, the products are: Sleek Phone, City Phone, Rice, Coscous, Beer, Women's shirt, Men's shirt.  We further imagine that an offline job has generated the embeddings or parameters $\bPsi,\brho$ already. These embeddings establish that there is a correlation such that users interested in the Sleek Phone are also interested in the City phone, users interested in Rice are also interested in Coscous, users interested in Beer have some interest in Rice and Coscous but it is not strong and finally users interested in the Men's Shirt are anti-correlated with users who are interested in the Women's Shirt. To make this concrete we assume that:

\[
  \bPsi =
  \left[ {\begin{array}{c}
    \bPsi_{\rm Sleek ~ Phone}\\
    \bPsi_{\rm City ~ Phone}\\
    \bPsi_{\rm Couscous}\\
    \bPsi_{\rm Rice}\\
    \bPsi_{\rm Beer}\\
    \bPsi_{\rm Female ~ Shirt}\\
    \bPsi_{\rm Male ~ Shirt}
  \end{array} } \right]
=
  \left[ {\begin{array}{ccccc}
  .9 & 0.05 & 0 & 0.05 & 0 \\
  1 & 0 & 0 & 0 & 0\\
  0 & .95 & 0 & 0.1 & 0\\
  0 & 1 & 0 & 0 & 0\\
  0 & 0.2 & .7 & 0 & 0\\
  0 & 0 & 0 & 1 & -1\\
  0 & 0 & 0 & -1 & 1
 \end{array} } \right]
\]

\[
  \brho = \boldsymbol{0}
\]

\noindent
The fact that most entries in $\bPsi$ are positive simplifies the discussion as it means that $\bomega$ high implies interest, The only two negative values are used to model a negative correlation between female and male shirts.

An interesting facet of this model is that it is not trivial to establish which is the most popular product simply by examining the parameters, while $\brho$ does reflect popularity it is also affected by $\bPsi$ in complex ways. In this constructed example we are able to label the five components of $\bomega$ as phones, grains, drinks, women's clothes and men's clothes.  

We now consider how different user histories affect \newline $p(\bomega|v_{u,1}, .., v_{u,T_u}, \bPsi, \brho)$. Approximation of this quantity can be made accurately and easily using the Stan probabilistic programming language \cite{stan2018}, although later we will show that excellent performance can also be obtained from using variational methods, that are viable to scale to real world recommender systems, and with comparative cost to methods such as Recurrent Neural Networks.

\begin{figure}
  \centering
      \begin{subfigure}[b]{0.24\textwidth}            
              \includegraphics[width=0.9\textwidth]{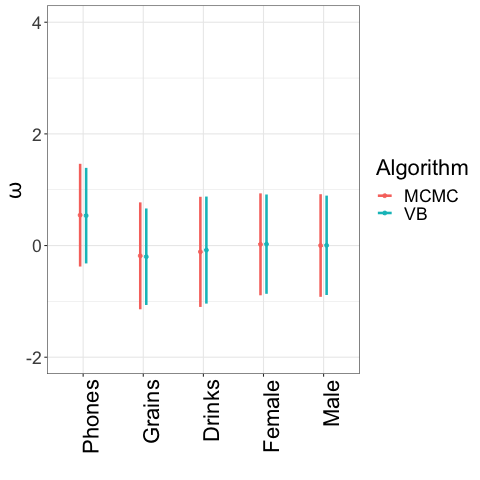}
              \label{fig:SRl}
      \end{subfigure}%
      \begin{subfigure}[b]{0.24\textwidth}
              \centering
              \includegraphics[width=0.9\textwidth]{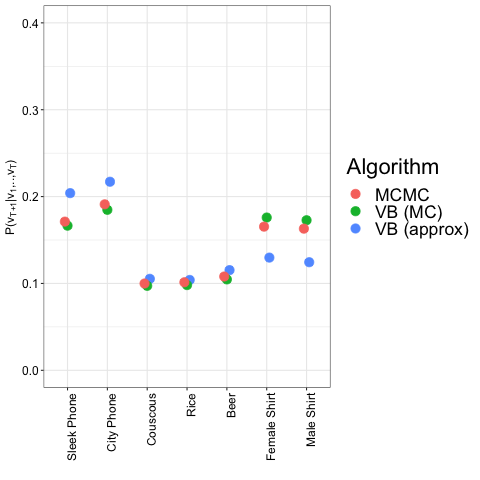}
              \label{fig:D-Imager}
      \end{subfigure}
  \caption{User representation (left) and next item prediction for a user with one sleek phone in their history}
  \label{fig:TOF1}
  \begin{subfigure}[b]{0.24\textwidth}            
        \includegraphics[width=0.9\textwidth]{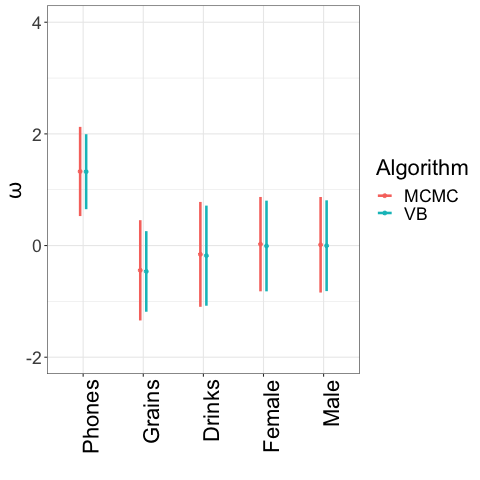}
        \label{fig:SRl}
\end{subfigure}%
\begin{subfigure}[b]{0.24\textwidth}
        \centering
        \includegraphics[width=0.9\textwidth]{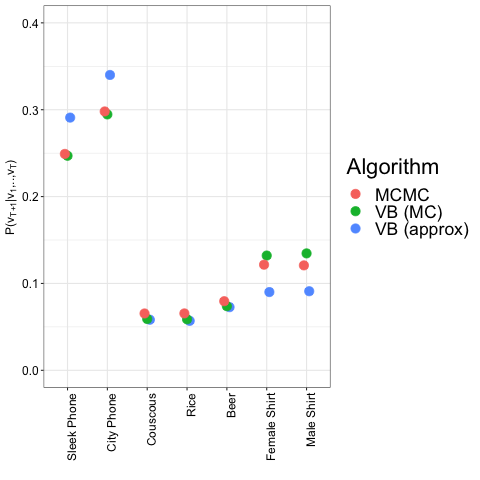}
        \label{fig:D-Imager}
\end{subfigure}
\caption{User representation (left) and next item prediction for a user with one sleek phone and two city phones in their history}\label{fig:TOF2}
\begin{subfigure}[b]{0.24\textwidth}            
  \includegraphics[width=0.9\textwidth]{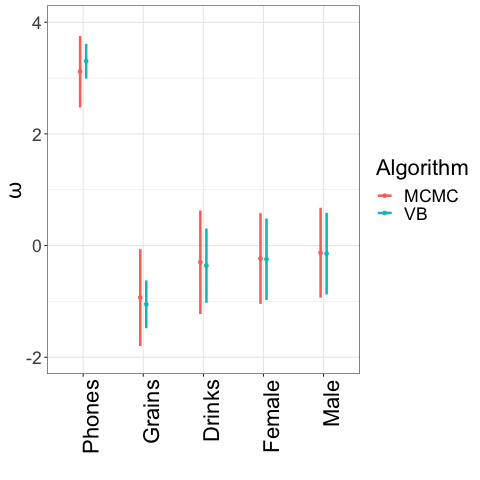}
  \label{fig:SRl}
\end{subfigure}%
\begin{subfigure}[b]{0.24\textwidth}
  \centering
  \includegraphics[width=0.9\textwidth]{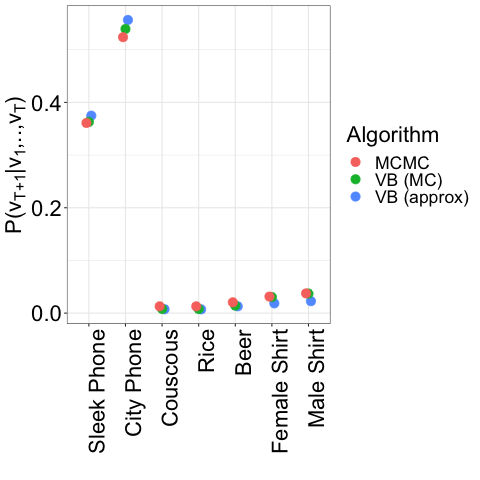}
  \label{fig:D-Imager}
\end{subfigure}
\caption{User representation (left) and next item prediction for a user with one sleek phone and twenty city phones in their history}\label{fig:TOF3}
\begin{subfigure}[b]{0.24\textwidth}            
  \includegraphics[width=0.9\textwidth]{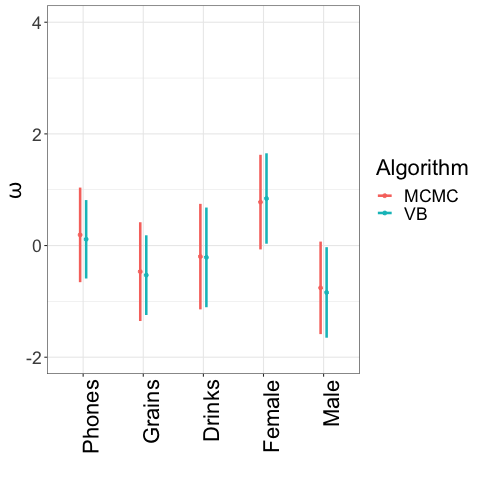}
  \label{fig:SRl}
\end{subfigure}%
\begin{subfigure}[b]{0.24\textwidth}
  \centering
  \includegraphics[width=0.9\textwidth]{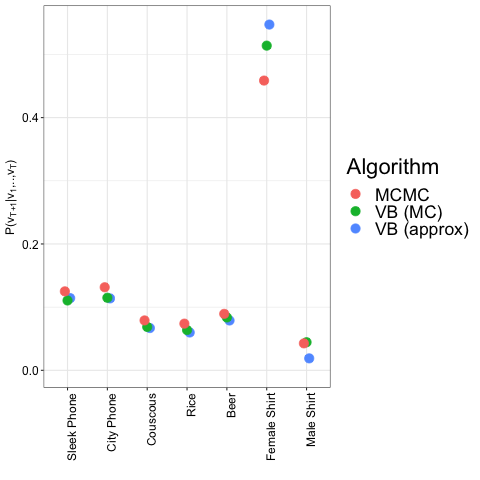}
  \label{fig:D-Imager}
\end{subfigure}
\caption{User representation (left) and next item prediction for a user with two female shirts and one sleek phone in their history}
\label{four}
\end{figure}

In Figure~\ref{fig:TOF1} - \ref{four} the intuitive behavior of this simple model is demonstrated. The results of three approximate methods are presented and shown to be in good agreement, we defer discussion of the approximation methods here except to note that we take Markov chain Monte Carlo to be the gold standard.

In Figure~\ref{fig:TOF1} we observe the case where a single Sleek Phone is observed in the user's history as a consequence of the short history there is significant uncertainty in the knowledge about the user, although the embedding reflects higher interest in the phone category; the next item prediction is high for both the Sleek Phone and the City Phone.  In Figure~\ref{fig:TOF2}  we observe the case where a single Sleek Phone is observed twice and the City Phone once in the user's history as a consequence there is  less uncertainty in the knowledge about the user, the next item prediction is higher for both the Sleek Phone and the City Phone.  In Figure~\ref{fig:TOF3}  the user has viewed the city phone twenty times and the Sleek phone just once, as a consequence the user's embedding shows a strong interest in phones with low uncertainty, the next item prediction is distributed among the two phones.  In Figure~\ref{four}  we observe a user who has observed a City Phone and a Women's Shirt, we see that the user's interests in phones and women's clothes are increased and their interest in men's clothing is decreased, indicating the negative entry in the embedding has the desired effect.  The next item predictions also reflects these preferences.

This simple model has shown a remarkable ability to summarize a user's interests and is able to reflect both strong or weak information about our knowledge of the user.  Having demonstrated the intuitive value of this model we now show how to estimate $\bPsi,\brho$, how to efficiently approximate $p(\bomega|v_{u,1}, .., v_{u,T_u}, \bPsi, \brho)$ such that it can be done online updating the user representation as the user acts in a dynamic way and finally how to do next item prediction which may be a proxy to the recommendation task.

\section{Related literature}
\label{sec:lit-review}

\subsection{Scalable Variational Approximations}

Two approaches for scalable Bayesian inference focus on approximating a posterior on a fixed dimensional parameters space rather than the latent variable case we care about as such they are not appropriate for our case \cite{kucukelbir2017automatic}, \cite{ranganath2013black}.

Under a conditional independence assumption it is often possible to reduce the variational Expectation Maximization algorithm to a finite sum fixed point iteration, where the finite sum is over the data plus another term associated with the prior and the entropy.  This formulation rather directly allows the  Robbins Monro stochastic approximation algorithm \cite{robbins1951stochastic}  and has been effective in complete data exponential family models \cite{hoffman2013stochastic}.  While this algorithm does indeed apply to simple versions of our model the "M-Step" for estimating the embeddings would require un-feasibly large matrix inverses.

\subsection{Latent Variable Models}

Our model is a special case of \cite{liang2018variational} (also see \cite{rezende2014stochastic,lafferty2006correlated}) which has stronger analytical properties including an analytical bound and EM algorithm which we can exploit both to gain computational advantages and to highlight similarities with other methods.  

An interesting suggestion made in \cite{liang2018variational} is the use of reducing the contribution of the Kullback Leibler component of the lower bound e.g. by multiplying this by a value lower than one e.g. one half.  They justify this with a combination of empirical results and by interpreting the model as containing a construction error and regularization.  In this paper we are primarily focused upon producing a user representation, to multiply the KL component by a half would have the effect of ``squaring the likelihood'' i.e. double counting the data resulting in artificially reduced uncertainties on the user representation.  As we are primarily interested in producing a user representation we do not pursue that method here, although we do acknowledge the excellent empirical results they present.

We use a semi-Bayesian or latent variable framework integrating the latent variable but estimating the parameters.  There is a literature discussing the improved statistical properties of this procedure, for see theoretical arguments given in \cite{welling2008deterministic} for a demonstration of empirical performance see \cite{NIPS2011_4273}.  A critical observation is that for a traditional matrix factorization the parameter space grows with the number of users, this makes traditional statistical notions such as convergence difficulty and indeed means that if a new user arrives a fit must be done before a prediction can be done.

In contrast if one of the matrices is integrated then the model becomes fixed dimensional then the dimensionality of the model is fixed and traditional statistical notions such as convergence again become relevant.

\subsection{Word2Vec and Prod2Vec}

The skipgram model and skipgram with negative sampling, collectively known as word2vec \cite{mikolov2013efficient}, caused a sensation in both the natural language and recommender systems community \cite{gunawardana2009unified}. If we define the event matrix:

\[
  \bC_{u,p} = \sum_{t=1}^{T_u} \boldsymbol{1}\{ v_t=p \}.
\]

\noindent
We then let $\tilde{\bC}_{u,p} = {\bC}_{u,p} > 0$, i.e.  is thresholded to either 0 or 1, then word2vec operates on the co-event matrix: 
$\tilde{\bC}^T\tilde{\bC}$, often the rows and columns are refereed to as target and context.  This matrix is of size $P \times P$ which while large is often much smaller than $\bC$ which is $P \times U$ where $U$ is the number of users, so operating on this matrix is more efficient computationally.  There are however disadvantages in modeling $\tilde{\bC}^T\tilde{\bC}$ directly.  One being that $\tilde{\bC}^T\tilde{\bC}$ has some quite subtle relationships e.g. some values of $\tilde{\bC}^T\tilde{\bC}$ are impossible.  An example of a valid matrix is:

\[
  \tilde{\bC}^T\tilde{\bC}= \left(\begin{matrix}
1 & 1 & 0\\
1 & 1 & 0\\
0 & 0 & 1\\
\end{matrix}
\right).
\]

\noindent
This matrix is consistent with two user sessions, the first session visited product 1 and 2 and the second session visited product 3.  Now consider:

\[
  \tilde{\bC}^T\tilde{\bC}\ne 
\left(
\begin{matrix}
1 & 1 & 1\\
1 & 1 & 0\\
1 & 0 & 1\\
\end{matrix}
\right)
\]

\noindent
This matrix is inconsistent with any $\tilde{\bC}$ containing only positive counts.  Intuitively we can see this by noting that the diagonal implies that each product has been observed as associated with one user each.  The first row (or by symmetry the first column) says that product 1, 2 and 3 all occur together.  The only way we can have each product viewed exactly once and all occurring together is for all entries to be associated yet we have $[\tilde{\bC}^T\tilde{\bC}]_{3,2}=[\tilde{\bC}^T\tilde{\bC}]_{2,3}=0$ which is inconsistent.  The skipgram model suggests modeling the rows of $\bC$ as multinomial draws, which gives positive probability to events that cannot happen, given the complexity of the constraints on $\tilde{\bC}$ it is difficult to see how this can be avoided except by modeling $\tilde{\bC}$ or even $\bC$ directly.

A further contribution in \cite{mikolov2013efficient} was a negative sampling heuristic, which allowed these methods to scale to very large numbers of categories by avoiding large summations over every iteration.   However the meaning of negative sampling remains unclear and it complicates producing probabilistic algorithms.  For example within these classes of algorithms there is a tuning parameter to decide how many negative examples to generated.  Of course increasing the amount of (artificially)  generated data will (artificially) reduce uncertainty on parameter estimates, while there have been attempts at a Bayesian skipgram model \cite{barkan2017bayesian} it is difficult to see how any method employing this heuristic can correctly control the uncertainties that they compute.

It is interesting to reflect on the widespread successful use of these methods.  The heuristics employed do not make it easy to make a complete comparison with our method, but we can make a few comments.
If the underlying model of $\bC$ was Gaussian (this cannot be true as it has support only on natural numbers) then 
$\bC^T\bC$ would be the scatter matrix which along with the mean gives the sufficient statistic of a Gaussian distribution.  Taking the eigenvalue decomposition of this would result in  principal component analysis (without the usual subtracting of the means step) and there is the well known result that PCA can be computed either by an eigenvalue decomposition of $\bC^T\bC$ or the singular value decomposition of $\bC$, which loosely accounting for the changes of support and the integration is what our method achieves.
Loosely we can view our latent variable model also in sense i.e. estimating the covariance as $\bPsi \bPsi^T$ or doing a matrix factorization of the form $\bC_u \propto \exp( \bPsi \bomega_u + \brho)$. The use of dot products between embeddings can be viewed as covariances and the cosine distances as correlations.

The non-probabilistic nature of word2vec poses problems that are typically dealt with using heuristics such as using these embeddings as features in the ``feature engineering approach'' several questions are difficult to resolve e.g. How do you do next item prediction (combining popularity with the associated embeddings)?  How do you do recommendation?  How do you combine several items of history into the a fixed dimensional user state?

\subsection{RNN Session Based Recommendation}

In the session based recommender system literature, there is a significant literature applying RNNs to the recommendation problem in this case like us they apply the model directly to $\bv$.  The RNN is a more flexible model able to capture more sophisticated sequences e.g. if a shopper transitions from being interested to complimentary products after a purchase event.  This extra flexibility is powerful, but also require more data to identify these effects.  

In contrast the latent variance model we introduce is effectively a low rank Gaussian prior on a categorical variable as such up to the capacity of the model the law of large numbers would apply i.e. if a user had a long enough history and the embedding size $K$ was greater than equal to the number of products $P$ then the next item prediction would converge to the empirical history due to the law of large numbers\cite{de1937prevision}.  In contrast the RNN does not a priori incorporate the law of large numbers it is a flexible sequential model and if the law of large numbers holds, as we might approximately expect, then the RNN will need to see more data to recognize this.  It is an important remark that if the user embedding size is less than the number of products $K<P$ regardless of if an RNN or a latent factor model is used it is not possible for the next item prediction for a user to converge to their empirical history due to a lack of capacity.  The non-linear model of \cite{liang2018variational} also has the same limitation.
This low capacity is typically not a problem as user sequences are very short, but it does highlight that there are limitations introduced by using small embedding sizes i.e. the ability to distinguish users in subtly different products may be lost; of course the advantage is vastly improved tractability.  The stronger assumptions of the latent variable method suggest its realm of applicability is when those assumptions are true, or they are approximately true and there is insufficient data to learn a higher capacity model such as an RNN.

\section{Approximate Inference}
\label{sec:aprox-inf}

In previous sections we discussed the model and showed it has intuitively reasonable properties.  In this section we show (i) how to learn the embeddings $\bPsi,\brho$ and (ii) how, at deployment, to make predictions by approximating the posterior over a user's representation i.e. how to compute $p(\bomega|v_{u,1},..v_{u,T_u})$ in real time.

\subsection{Optimizing the lower bound}

In order to make this method practically usable we need two components: firstly to be able to estimate $\bPsi,\brho$ efficiently and secondly we need to be able to rapidly produce and update user embeddings based on a user's activity. To solve both parts of the problem, we will employ variational approximations. Variational approximations work by turning integration problems into optimization problems.

The model we introduce has the form:

\begin{align*}
  \log ~  & p(v_1,..,v_T,\bomega_u|\bPsi)  = \left( \sum_t^T \bPsi_{v_t}
            \bomega_u + \brho_{v_t}\right) \\
&  -T \log\{ \sum_p^P \exp( \bPsi_p \bomega_u  +\brho_p)\}  - \frac{K}{2} \log( 2 \pi )-
                                       \frac{1}{2}  \bomega_{u}^T
                                       \bomega_{u} 
\end{align*}

\noindent
If we use a normal distribution $\bomega \sim \mathcal{N}(\bmu_q,\bSigma_q)$, then variational bound has the form:

\begin{align*}
& \mathcal{L} = \E_{q(\bomega)}[\log ~   p(v_1,..,v_T,\bomega_u|\bPsi) -\log q(\bomega)] = \\
&\left( \sum_t^T \bPsi_{v_t}
            \bmu_q + \brho_{v_t}\right)  -T \E_{q(\bomega)} [\log \{\sum_p^P \exp( \bPsi_p \bomega_u  +\brho_p)\}]\\
&  - \frac{K}{2} \log( 2 \pi )-  \frac{1}{2}  \{ \bmu_q^T \bmu_q +
  {\rm trace} (\bSigma_q)  \} + \frac{1}{2} \log |2 \pi e \bSigma_q |
\end{align*}

We see that there is a problematic term associated with the denominator of the softmax.  We consider two possible computational approaches to this the Bouchard bound \cite{bouchard2007efficient} and the re-parameterization trick \cite{kingma2013auto}.

\subsubsection{Bouchard Bound}
The Bouchard bound introduces a further approximation and additional variational parameters $a,\xi$ but produces an analytical bound:

\begin{align*}
& \mathcal{L} \ge  \mathcal{L}_{\rm Bouch}  = \left( \sum_t^T \bPsi_{v_t} \bmu_q  + \brho_{v_t} \right) \\
  & -T [ 
  a + \sum_p^P \frac{\bPsi_p \bmu_q +\brho_p - a - \xi_p}{2} \\
  & + \lambda_{\rm JJ}(\xi_p) 
    \{(\bPsi_p \bmu_q+\brho_p-a)^2 + \bPsi_p \bSigma_q \bPsi_p^T  - \xi_p^2  \} + \log(1 + e^{\xi_p})
  ]\\  
  &  - \frac{K}{2} \log( 2 \pi )-  \frac{1}{2}  \{ \bmu_q^T \bmu_q +
  {\rm trace} (\bSigma_q)  \} + \frac{1}{2} \log |2 \pi e \bSigma_q |.
\end{align*}

\noindent
Where $\lambda_{\rm JJ}(\cdot)$ is the Jaakola and Jordan function \cite{jaakkola1997variational}:

  \[
  \lambda_{\rm JJ}(\xi) = \frac{1}{2\xi} \left( \frac{1}{1+e^{-\xi}} - \frac{1}{2} \right).
  \]

\noindent
  The bound may be optimized using the following variational EM algorithm which enjoys the coordinate descent properties of an EM algorithm guaranteeing the bound will tighten at each iteration. The algorithm here is the \emph{dual} of the one presented in \cite{bouchard2007efficient} as we assume the embedding $\bPsi$ is fixed and $\bomega$ is updated where the algorithm they present does the opposite.  The EM algorithm consists of cycling the following update equations:

\[
\bSigma_q^{-1} = I_k + 2 T \sum_p \lambda_{\rm JJ}(\xi_p)\bPsi_p^T \bPsi_p,
\]

\[
\bmu_q = \bSigma_q  \left(  ( \sum_t^T \bPsi_{v_t}^T) - T \left[ \sum_p^P
  \{ \frac{1}{2} + 2(\brho_p-a) \lambda_{\rm JJ}(\xi_p) \} \bPsi_p^T  \right] \right),
\]

\[
a = \frac{-1+\frac{P}{2} + \sum_p  2 \lambda_{\rm JJ}(\xi_p) (\bPsi_p \bmu_q + \brho_p) }{ 2 \sum_p \lambda_{\rm JJ}(\xi_p) },
\]

\[
\xi_{p} = \sqrt{\bPsi_p \bSigma_q \bPsi_p^T + (\bPsi_p \bmu_q + \brho_p-a)^2 }.
\]

There are other variational bounds that may be considered for this problem most notably the tilted bound \cite{knowles2011non}.  Even though the Bouchard bound is loose compared to the tilted bound, it does enjoy the availability of an EM algorithm which enjoys the stability properties of a coordinate descent algorithm. In the case of the tilted bound the known fixed point algorithms are not guaranteed to be stable and are not always stable in practice \cite{nolan2017accurate,rohde2016semiparametric} so extra methods such as line searches would need to be considered.  We do not further consider alternative bounds.

The computational cost of this algorithm depends on the number of products $P$ linearly and the embedding size $K$ cubicly, if $P$ and $K$ are modest it can take less than a second making it potentially deployable at prediction time.  In practice we found the cost of large $P$ might be prohibitive due to the sums over all $P$ embeddings, in these cases a variational auto-encode described in the next section, is to be preferred.

\subsubsection{Re-parameterization Trick}

The second approach to computing expectations with respect to the denominator of the softmax is to use the re-parameterization trick \cite{kingma2013auto}, which allows us to take a  sample of $\bomega$ from the variational distribution and compute a noisy derivative of the lower bound.  Within each iteration we proceed by simulating:

\[
\bepsilon^{(s)} \sim \mathcal{N}(\boldsymbol{0}_K, \bI_K),
\]

\noindent
and then computing:

\[
\bomega^{(s)} = L_{\bSigma_q} \bepsilon^{(s)} + \bmu_q.
\]

\noindent
Where $\bL_{\bSigma_q} \bL_{\bSigma_q}^T = \bSigma_q$, we can then optimize the noisy lower bound:

\begin{align*}
  & \mathcal{L}_{MC} = \\
  &\left( \sum_t^T \bPsi_{v_t}
              \bmu_q + \brho_{v_t}\right)  -T \log[ \sum_p^P \exp\{ \bPsi_p (L_{\bSigma_q} \bepsilon^{(s)} + \bmu_q  )  +\brho_p\}]\\
  &  - \frac{K}{2} \log( 2 \pi )-  \frac{1}{2}  \{ \bmu_q^T \bmu_q +
    {\rm trace} (\bSigma_q)  \} + \frac{1}{2} \log |2 \pi e \bSigma_q |.
\end{align*}  

\noindent
Often $\bSigma_q$ is taken to be diagonal which makes computing $\bL_{\bSigma_q}$ simply an element-wise square root.

\subsection{Latent variable size growing with data}

A naive application of the algorithm discussed so far would have the number of variational parameters $\bmu_q,\bSigma_q$ or $\bmu_q,\bSigma_q, \bxi, a$ for the Bouchard bound growing with the number of parameters. We propose to limit the number of parameters by the use of a variational auto-encoder \cite{kingma2013auto}.  This involves using a flexible function and optimizing it to do the job of the EM algorithm i.e.

\[
\bmu_q, ~ \bSigma_q = f_\Xi(v_1,...v_{T}),
\]

\noindent
or in the case of the Bouchard bound:

\[
\bmu_q, ~ \bSigma_q, ~ a, ~ \xi = f_\Xi^{\rm Bouch} (v_1,...v_{T}).
\]

\noindent
Where any function e.g. a deep net can be used for $f_\Xi(\cdot)$ and $f_\Xi^{\rm Bouch}(\cdot)$.

It is common to use the re-parameterization trick and an auto-encoder in combination although this is not necessary.  The choice between the two hinges on accepting Monte Carlo error or using a looser, but analytical bound.

\subsection{Next Item Prediction}
Finally and perhaps surprisingly the predictive distribution required to do next item prediction is also not trivial in this case, i.e. approximating:

\begin{align*}
&  p(v_{u,T+1}|v_{u,1},..,v_{u,T}) \\
&  = \int  p(v_{u,T+1}|\bomega,\bPsi,\brho) p(\bomega|v_{u,1},.v_{u,T})  d \bomega_u
\end{align*}

\noindent
is not trivial even if $p(\bomega|v_{u,1},..v_{u,T_u})$ is approximated with a Gaussian distribution $\bomega_u|v_1,..v_T \sim \mathcal{N}(\bmu_q,\bSigma_q)$.  We are interested in computing:

\[
 p(v_{n+1}|v_1,...v_n) \approx \E_{q(\bomega)} \left[ \frac{\exp(\bPsi_v \bomega + \brho) }{\sum_{v'} \exp(\bPsi_{v'} \bomega + \brho)  }\right].
\]

\noindent
We considered using a Monte Carlo based approximation, first by drawing $S$ samples:
\[
\bomega^{(s)}  \sim \mathcal{N}(\bmu_q, \bSigma_q),
\]

\[
  p(v_{n+1}|v_1,...v_n) \approx \frac{1}{S} \sum_s^S \frac{\exp(\bPsi_v \bomega^{(s)} + \brho) }{\sum_{v'} \exp(\bPsi_{v'} \bomega^{(s)} + \brho)  },
\]

\noindent
as well as using a number of fast approximations such as:

\[
  p(v_{n+1}|v_1,...v_n) \approx  \frac{\exp(\bPsi_v \bmu_q + \brho) }{\sum_{v'} \exp(\bPsi_{v'} \bmu_q + \brho)  }.
\]

\noindent
while we investigated more complex approximations (such as normalizing the exponential of the lower bound) we did not find they helped in practice, the two VB approximations shown in Figure~\ref{fig:TOF1} - \ref{four} and denoted (MC) and (approx) are the Monte Carlo and mean approximations respectively.

\section{Experimental Setup}
\label{sec:experimental-setup}

We demonstrate that our method produces useful user representations on next item prediction using the RecoGym simulation environment \cite{rohde2018recogym}. RecoGym is a framework for simulating a recommender system and enables the simulation of AB tests although here we simply use it to create organic sequences of item views and test the model's ability to do next item prediction, this allows us to compute the same metrics as on standard offline datasets. We also present results upon the YooChoose dataset \cite{ben2015recsys}. We split both the datasets into train and test so that sessions reside entirely in one of the two groups.  We fit the model to the training set, we then evaluate by providing the model $v_1,..v_{{T_u}-1}$ events and testing the model's ability to predict $v_{T_{u}}$.

\subsection{Implementation Details}

All the models, including the relevant baselines, have been implemented using the PyTorch automatic differentiation package in Python \cite{paszke2017automatic}. All models are updated via the use of Stochastic Gradient Descent (SGD), specifically the RMSProp variant. We set the learning rate to 0.001 and tune the other hyper-parameters, including L2 regularization, for each dataset based upon a validation set. The dataset specific hyper-parameter values are reported in Section \ref{sec:results} with the relevant results. 

\subsection{Performance Metrics}

The various models are evaluated using recall at K (RC@K) and truncated discounted cumulative gain at K (DCG@K), which are defined below.

Let $r_k$ be the $k$th highest value of $p(\bomega_{v_{T_{u}}}|v_1,..v_{{T_u}-1})$. For all results presented in this paper, we set K to five.

\begin{align*}
  {\rm RC@K} =\begin{cases}
    1, & \text{if $v_{T_u} \in \{r_1,...,r_K\}$}.\\
    0, & \text{otherwise}.
  \end{cases}
\end{align*}

\begin{align*}
  {\rm DCG@K} =
    \sum_i \frac{2^{r_i \boldsymbol{1}\{v_{T_u}\in \{r_1,...,r_K\} \} }-1 }{\log i+1}.
\end{align*}

\noindent
We compute the average of these quantities over all sessions in the test set.

\subsection{Latent Variable Inference}

\sloppy


We consider three alternative methods for training the model:

\begin{itemize}
  \item \textbf{Bouch/AE} - A linear variational auto-encoder using the Bouchard bound.
  \item \textbf{RT/AE} - Using the re-parameterization trick with the Bouchard bound.
  \item \textbf{RT/Deep AE} - A deep auto-encoder again using the re-parameterization trick. The deep auto-encoder consists of mapping an input of size P to three linear rectifier layers of K units each.  We encountered numerical problems using the Bouchard bound with a deep auto-encoder. 
\end{itemize}

When we update the posterior over a user's latent variable representation at test time, we assess both using the auto-encoder denoted AE and using the 100 iterations of the EM algorithm denoted EM in the results.

When we compute next item predictions we consider both using a 100 sample Monte Carlo approximation denoted MC and just taking the mean as a point estimate denoted mean it uses only $\bmu_q$ (and correspondingly ignores $\bSigma_q$).

\subsection{Baselines}

To demonstrate the effectiveness of our approach, we present results from the following baseline approaches:

\subsubsection{Popularity}

Item popularity provides no personalization, but is nonetheless a strong baselines for certain recommendation tasks.

\subsubsection{Item KNN}

Item K Nearest Neighbors (KNN) involves computing the correlation matrix of the sample data adding the identity to prevent division by zero and then using these correlations as recommendations based on a user's most recent historical item.  The limitations of this technique is that it ignores item popularity and multiple items in the user's history, but despite these limitations it is often a strong baseline.

\subsubsection{Recurrent Neural Network}

For this baseline, we make use of a recurrent neural network to learn a user representation by predicting the next item in the session. The model architecture we employ is similar to that of \cite{hidasi2015session}, in that we feed the output from an embedding layer into a Gated Recurrent Unit (GRU) \cite{cho2014learning} with 64 hidden units to learn the temporal dynamics of the user's session. The output from the GRU is then passed through a final softmax layer which gives the probability of the next item in the sequence. The network is trained to minimize the categorical cross-entropy over the training sessions via RMSProp.

\section{Results}
\label{sec:results}

\begin{table}
\begin{tabular}{lllrr}
  \toprule
  \textbf{Train}  & \textbf{Online}  & \textbf{Online}  &   \textbf{RC@5} &  \textbf{DCG@5} \\
  \textbf{Algorithm} &  \textbf{Latent} & \textbf{Next Item} &    &   \\
\midrule
\midrule
  Pop &          &        &  0.456 &   0.440  \T \\
  ItemKNN &          &        &  0.461 &   0.492 \\
        RNN &          &        &  0.620 &   0.646 \\
        \midrule
        Bouch/AE &          AE &        MC &  0.712 &   0.796 \\
        Bouch/AE &          AE &      mean &  0.712 &   0.777 \\
        Bouch/AE &          EM &        MC &  0.738 &   0.796 \\
        Bouch/AE &          EM &      mean &  \textbf{0.748} &   0.796 \\    
       RT/AE &          AE &        MC &  0.707 &   \textbf{0.802} \\
       RT/AE &          AE &      mean &  0.697 &   0.784 \\
       RT/AE &          EM &        MC &  0.738 &   \textbf{0.802} \\
       RT/AE &          EM &      mean &  0.733 &   \textbf{0.802} \\
   RT/Deep AE &          AE &        MC &  0.697 &   0.785 \\
   RT/Deep AE &          AE &      mean &  0.717 &   0.775 \\
   RT/Deep AE &          EM &        MC &  0.733 &   0.785 \\
   RT/Deep AE &          EM &      mean &  0.733 &   0.787 \\
  \bottomrule
  \end{tabular}
  \caption{Results on the testset for all approaches on the RecoGym dataset with 20 products. For both metrics, a higher value is better.}
  \vskip -15pt
  \label{recogym20}  
  \end{table}

\subsection{RecoGym}

For our first experiment we use the RecoGym simulator with 20 products and $\sigma_\omega=0$ i.e. a static user state. With this we generate a training set of 100 sessions and a test set of 1000 sessions, this results in 17161 and 176804 events for train and test respectively. The latent variable algorithms were all trained using 5000 epochs with the RMSProp algorithm and an embedding dimension of 10. The RNN was trained for 5000 epochs, with the same embedding size and again the RMSProp algorithm was used in all cases. The results from this are presented in Table \ref{recogym20}, which show the Bouchard method of training using the EM algorithm for predicting latent variables and Monte Carlo for predicting the next item was the best performing algorithm on the RC@5 metric, RT/AE performed slightly better on on the DCG@5 metric using either the EM algorithm or the auto-encoder with Monte Carlo.

\begin{table}
\begin{tabular}{lllrr}
  \toprule
  \textbf{Train}  & \textbf{Online}  & \textbf{Online}  &   \textbf{RC@5} &  \textbf{DCG@5} \\
  \textbf{Algorithm} &  \textbf{Latent} & \textbf{Next Item} &    &   \\
\midrule
\midrule
Pop &          &        &  0.020 &   0.016 \\
ItemKNN &          &        &  0.020 &   0.024 \\
RNN &          &        &  0.035 &   0.033 \\
\midrule
Bouch/AE &          AE &        MC &  0.082 &   0.128 \\
Bouch/AE &          AE &      mean &  0.082 &   0.079 \\
Bouch/AE &          EM &        MC &  \textbf{0.117} &   0.128 \\
Bouch/AE &          EM &      mean &  \textbf{0.117} &   \textbf{0.130}  \\
RT/AE &          AE &        MC &  0.061 &   0.047 \\
RT/AE &          AE &      mean &  0.056 &   0.059 \\
RT/AE &          EM &        MC &  0.051 &   0.047 \\
RT/AE &          EM &      mean &  0.051 &   0.047 \\
RT/Deep AE &          AE &        MC &  0.090 &   0.105 \\
RT/Deep AE &          AE &      mean &  0.080 &   0.068 \\
RT/Deep AE &          EM &        MC &  0.090 &   0.105 \\
RT/Deep AE &          EM &      mean &  0.090 &   0.106 \\
\bottomrule 
  \end{tabular}
  \caption{Results on the testset for all approaches on the RecoGym dataset with 2000 products. For both metrics, a higher value is better.}
  \vskip -15pt
  \label{recogym2000}  
  \end{table}

For our second experiment we use the RecoGym simulator with 2000 products and $\sigma_\omega=0$, i.e. a static user state, we generate a training set of 100 sessions and a test set of 100 sessions, this results in 21852 and 19533 events for train and test respectively.  The latent variable algorithms were all trained using 15000 epochs using the RMSProp algorithm, the embedding size was set to 10.  The RNN was trained with K=200 for 5000 epochs (it performed slightly worse with a training run of 25000). The results are shown in Table \ref{recogym2000}, again the Bouchard method of training using the EM algorithm for predicting latent variables and Monte Carlo for predicting the next item was the best performing algorithm on the RC@5 and DCG@5 metrics.  


\subsection{YooChoose}

\begin{table}
\begin{tabular}{lllrr}
  \toprule
  \textbf{Train}  & \textbf{Online}  & \textbf{Online}  &   \textbf{RC@5} &  \textbf{DCG@5} \\
  \textbf{Algorithm} &  \textbf{Latent} & \textbf{Next Item} &    &   \\
\midrule
\midrule
       Pop &          &        &   0.143 &   0.147 \\
    ItemKNN &          &        &   \textbf{0.804} &   \textbf{0.921} \\
        RNN &          &        &   0.690 &   0.781 \\
        \midrule
        Bouch/AE &          AE &        MC &   0.433 &   0.420 \\
        Bouch/AE &          AE &      mean &   0.451 &   0.562 \\
        Bouch/AE &          EM &        MC &   0.386 &   0.420 \\
        Bouch/AE &          EM &      mean &   0.429 &   0.497  \\           
        RT/AE &          AE &        MC &   0.495 &   0.731 \\
        RT/AE &          AE &      mean &   0.616 &   0.658 \\
        RT/AE &          EM &        MC &   0.693 &   0.731 \\
        RT/AE &          EM &      mean &   0.707 &   0.768 \\
        RT/Deep AE &          AE &        MC &   0.751 &   0.868 \\
        RT/Deep AE &          AE &      mean &   0.771 &   0.876 \\
        RT/Deep AE &          EM &        MC &   0.772 &   0.868 \\
        RT/Deep AE &          EM &      mean &   0.775 &   0.873 \\   
     \bottomrule
  \end{tabular}
  \caption{Results on the testset for all approaches on the YooChoose dataset with 100 products. For both metrics, a higher value is better.}
  \vskip -20pt
  \label{yc100}  
  \end{table}

For our third experiment we use the YooChoose dataset filtered to the most popular 100 products. This is a strong filter of YooChoose 60000 products, but allows for effective experimentation and still results in 2905816 events and 28286 events for the training and test set respectively.  The deep auto-encoder latent variable algorithms was  trained for 100 epochs, the linear Bouchard auto-encoder and re-parameterization trick auto-encoder were trained for 100 epochs, the RNN was trained for a single epoch and had an embedding size of 20, longer training runs were observed to cause overfitting and reduced performance.  All latent variables are trained using a full rank model i.e $K=100$.  The results are shown in Figure \ref{yc100}, in this case the ItemKNN model performs best on both metrics, the deep auto-encoder trained using the re-parameterization trick performs slightly worse, the best performing setups involve predicting using the mean method there was very little difference between predicting with the EM algorithm and with the auto-encoder on this data set.  


The ItemKNN baseline turned out to be very strong.  This is most likely due to the fact that we filtered the dataset to just 100 popular products allowing full rank covariance estimation.  The latent variable model also operating at full rank was unable to perform quite as well.  Another notable difference in the two methods is that ItemKNN just looks at the most recent event where the latent variable session model combines all history.  If the most recent event contains more relevant history this may advantage ItemKNN.  

\subsection{Interpretation of Results}

The model we present is very closely aligned to the internal model in the RecoGym simulator hence the strong performance here of all the variants of our model.  It is perhaps surprising that next item prediction using just the posterior mean performed similarly well to the Monte Carlo approach.  The value gained by the EM algorithm was also marginal.  Given the ability of an RNN to model very complex data such as language it is perhaps unsurprising that it performs poorly on the RecoGym 2000 product dataset given a relatively small sample.

For the YooChoose 100 product dataset the ItemKNN algorithm proved to be very effective.  The Deep AE was the closest performing with the EM MC variant being the best by a small margin.  The fact that the Deep AE performs the best and the linear auto-encoders improve substantially when using the EM algorithm both suggest that a linear auto-encoder is not sufficient for this problem.

\section{Conclusion}
\label{sec:conclusion}

Recommender systems are increasingly using embeddings to represent items. A user's session on the recommender system then will involve interactions with many of these items.  We have demonstrated an elegant algorithm for taking a user's history of varying length and summarizing it  with a posterior distribution over a user embedding which has the same dimension as the product embedding.  Sensible behavior such as higher uncertainty when the user has a short history and lower uncertainty when the user has a longer history are features of this model formulation.  We have demonstrated how it is possible to train the model to produce item embedding using a variational auto-encoder either with the re-parameterization technique or using the Bouchard bound.  Similarly it is possible to is possible to rapidly convert a user history containing multiple items to a user embedding using a variational auto-encoder or using the EM algorithm (although the later is constrained to small numbers of products due to summations over large $P$).  

A complexity of latent variable methods is the need to do a numerical integration at prediction time.  The EM algorithm presented has excellent stability properties, but scales poorly when the number of items is in the tens of thousands.  There are several lines of interesting work that could speed up this evaluation.  Alternatively using already well understood techniques we could simply use a variational auto-encoder, which also produces rapid approximation of the integral.  


There are numerous possible extensions to the training algorithm. Training speed requires normalization of size $P$ which can be prohibitive, methods such as those outlined in \cite{ruiz2018augment} may be adaptable to this model. Finally the model can be incorporated to model time in a more sophisticated way and to consider the feedback to recommendations rather than be exclusively built for next item prediction.

\bibliographystyle{ACM-Reference-Format}
\bibliography{literature} 


\end{document}